\documentclass{ws-procs975x65}

\newcommand{\ffat}[1]{\mbox {\boldmath $#1$}}

\begin{document}

\title{Three-Nucleon scattering at Intermediate Energies}

\author{I. Fachruddin}
\address{Departemen Fisika, Universitas Indonesia, Depok 16424, Indonesia}

\author{Ch. Elster}
\address{Department of Physics and Astronomy, Ohio University, Athens, 
OH 34701, USA}  

\author{W. Gl\"ockle}
\address{Institut f\"ur Theoretische Physik II, Ruhr-Universit\"at Bochum, 
D-44780 Bochum, Germany}  

\maketitle

\abstracts{
By means of a technique, which does not employ partial wave (PW)
decompositions, the nucleon-deuteron break-up process is
evaluated in the Faddeev scheme, where only the leading order term of the
amplitude is considered. This technique is then applied to
calculate the semi-exclusive proton-deuteron break-up reaction $d(p,n)pp$ 
for proton laboratory energies $E_{lab}$ of a few hundred MeV. 
A comparison with PW calculations is performed 
at 197 MeV projectile energy. 
At the same energy rescattering processes, which are not 
included in the 3D calculations yet, are shown to be still important 
in the full Faddeev PW calculations, especially for the cross section 
and the analyzing power $A_y$. Next kinematical relativistic effects are 
investigated for projectile energies up to about 500 MeV. 
At the higher energies, those relativistic effects start not to be negligible, 
especially in the peak of the cross section.}


\section{Introduction}
In order to investigate the short distance behavior of 
three nucleon (3N) as well as nucleon-nucleon (NN)
forces it is necessary to consider 3N scattering at higher
energies of a few hundred MeV. The partial wave (PW) 
technique, which takes as basis states a specific number of 
angular momentum eigenstates, has a long history of being employed to solve Faddeev 
equations for 3N scattering [\refcite{faddeev}].  
However as the energy increases the number of contributing angular momenta
proliferates, leading to increased algorithmic difficulties as well as 
more tedious algebraic work. 
Therefore, a new and alternative approach is
needed, which is not based on PW basis. 

In momentum space the choice is to work directly with momentum vector states.
For three boson scattering this approach has been pioneered successfully carried
out in Refs.~[\refcite{3bosons3d}]. 
We develop our approach [\refcite{3n3d}], referred here as the 3D technique, 
for three nucleon scattering, allowing the use of realistic nuclear forces, 
since spin and isospin are also taken into account. 
Since the NN system serves as input for our 
3N calculations, we 
successfully applied the 3D technique to both NN scattering [\refcite{nn3d}] and 
the deuteron [\refcite{deut3d}]. 

We evaluate the nucleon-deuteron (Nd) break-up process using the Faddeev
scheme. However, we do not solve the full Faddeev equations but rather
concentrate on the first order term in the multiple scattering series 
given by them. Thus, we 
assume that at intermediate energies of a few hundred MeV
the leading term may sufficiently describe
the scattering process. We include relativistic effects in the kinematics to
investigate their size as function of energy. 
For our application we concentrate on the
semi-exclusive $d(p,n)pp$ reaction and calculate the spin averaged 
differential cross section, 
the neutron polarization, the proton analyzing power and the polarization 
transfer coefficients. There are in fact experimental data in the energy range 
up to about 500 MeV [\refcite{495}]-[\refcite{197}]. 
The 3D technique works with any potentials given in operator
form, not the PW projected ones. We employ the NN potentials  AV18 [\refcite{av18}] 
and Bonn-B [\refcite{bonnb}].


\section{The Nd Break-Up Amplitude}
The break-up process is given by the Nd break-up operator $U_0$ given by 
\begin{equation}
U_0 = (1 + P)TP ,
\end{equation}
where $P \equiv P_{12}P_{23} + P_{13}P_{23}$ is a permutation operator, 
$T$ stands for NN t-matrix. The term $TP$ is the leading term of the 3N transition
operator 
$T_F$ obeying Faddeev equation $T_F = TP + TG_0PT_F$, with
$G_0$ being the free three-nucleon propagator. The operator $U_0$ is 
fully anti-symmetrized and can be written as a sum of three terms in $TP$, 
the first of which is $U_0^{(1)}=TP$. The other terms 
$U_0^{(2)}=P_{12}P_{23}TP$ and $U_0^{(3)}=P_{12}P_{23}TP$ are related 
to $U_0^{(1)}$ by means of permutations. 

Without employing PW decompositions, except for the deuteron, the Nd break-up 
amplitude $U_0({\bf p},{\bf q})$ is  given in
Eq.~(\ref{thai1}) (see Ref.~[\refcite{3n3d}] for derivation), 
where $m_{si},\tau_i$ ($i = 1,2,3$) are the spins and 
isospins of the three nucleons in final state, $m_{s1}^0,\tau_1^0$ 
the spin and isospin of nucleon 1 acting as the projectile, $\Psi
_d^{M_d}$ the deuteron state, $M_d$ 
the projection of total angular momentum of the deuteron along an arbitrary 
z-axis, $\psi _{l}(\pi ') $ the deuteron wave function components (s and
d waves), $T_{\Lambda \Lambda '}^{\pi St}(p,\pi,\cos \theta ';E_p)$ the NN
t-matrix elements in a momentum-helicity basis [\refcite{nn3d}]
for given parity $\pi $, total spin $S$ and isospin $t$, final and
initial helicities $\Lambda $ and $\Lambda '$, and a center of mass energy 
$E_p$ of the 23-subsystem, $d^{S}_{\Lambda '\Lambda }(\theta )$'s being the
rotation matrices [\refcite{rose}]. The Jacobi 
momenta  ${\bf p}$ and ${\bf q}$ describe the
3N kinematics in the final state, 
where ${\bf p}$ is the relative momentum between nucleon 2 and 3, 
${\bf q}$ the relative momentum of nucleon 1 to the 23-subsystem, 
and ${\bf q}_0$ the relative momentum of the projectile to the deuteron. 

The operator $U_{0}({\bf p},{\bf q})$ given in Eq.~(\ref{thai1}) is derived within the 
framework of the nonrelativistic Faddeev scheme. By adopting the formulation 
given in  Ref.~[\refcite{fong}], relativistic kinematics is introduced.
Thus, we re-evaluate the Jacobi momenta, 
carry out corresponding Lorentz transformations to the two- and 
three-particle c.m. subsystems, and employ relativistic energy-momentum relations. 
To calculate the observables, a relativistic description of the cross
section is employed. 

\begin{eqnarray}
U_0({\bf p},{\bf q}) & \equiv & \left\langle {\bf p}{\bf q}m_{s1}m_{s2}m_{s3}\tau_1\tau_2\tau_3 \biggl| U_0 \biggr| {\bf q}_0m_{s1}^0\tau_1^0 \Psi _d^{M_d}\right\rangle \nonumber\\
& = & \frac{(-)^{\frac{1}{2}+\tau _{1}}}{4\sqrt{2}}\delta _{\tau _{2}+\tau _{3},\tau _{1}^0-\tau _{1}}\sum _{m_{s}'} e^{-i(\Lambda _{0}\phi _p -\Lambda _{0}'\phi _{\pi })} C\left( \frac{1}{2}\frac{1}{2}1;m_{s}'m_{s1}\right) \nonumber\\
 &  & \times \sum _{l}C\left( l11;M_{d}-m_{s}'-m_{s1},m_{s}'+m_{s1}\right) Y_{l,M_{d}-m_{s}'-m_{s1}}(\hat{{\ffat \pi }}')\psi _{l}(\pi ') \nonumber\\
&  & \times \sum _{S\pi t}\left( 1-\eta _{\pi }(-)^{S+t}\right) C\left( \frac{1}{2}\frac{1}{2}t;\tau _{2}\tau _{3}\right) C\left( \frac{1}{2}\frac{1}{2}t;\tau _{1}^0,-\tau _{1}\right) \nonumber\\
&  & \times C\left( \frac{1}{2}\frac{1}{2}S;m_{s2}m_{s3}\Lambda _{0}\right) C\left( \frac{1}{2}\frac{1}{2}S;m_{s1}^0m_{s}'\Lambda _{0}'\right) \nonumber\\
 &  & \times \sum _{\Lambda \Lambda '}d^{S}_{\Lambda _{0}\Lambda
 }(\theta _p)d^{S}_{\Lambda _{0}'\Lambda '}(\theta _{\pi })
 e^{i(\Lambda '\phi '-\Lambda \Omega )}T_{\Lambda \Lambda '}^{\pi
   St}(p,\pi,\cos \theta ';E_p) \nonumber\\
 &  & + U_0^{(2)}({\bf p},{\bf q}) + U_0^{(3)}({\bf p},{\bf q}) , \label{thai1}
\end{eqnarray}
with
\begin{eqnarray}
{\ffat \pi } & \equiv & \frac{1}{2}{\bf q}+{\bf q}_0 \qquad \qquad {\ffat \pi }'\equiv -{\bf q}-\frac{1}{2}{\bf q}_0 \\
\cos \theta ' & = & \cos \theta _p\cos \theta _{\pi }+\sin \theta _p\sin \theta _{\pi }\cos (\phi _p-\phi _{\pi })\\
e^{i(\Lambda '\phi '-\Lambda \Omega )} & = & \frac{\sum ^{S}_{N=-S}e^{iN(\phi _p-\phi _{\pi })}d^{S}_{N\Lambda }(\theta _p)d^{S}_{N\Lambda '}(\theta _{\pi })}{d^{S}_{\Lambda '\Lambda }(\theta ')} .
\end{eqnarray}

\section{The Semi-exclusive Proton-Deuteron Break-Up Reaction}
Our formulation is applied to the semi-exclusive proton-deuteron (pd) break-up
reaction $d(p,n)pp$. We calculate the spin averaged differential cross section, 
the neutron polarization, the proton analyzing power and the polarization 
transfer coefficients for proton laboratory energies $E_{lab}$ up to about 500
MeV. 

First, we perform comparisons with calculations based on the well 
established PW technique, which also include only the leading term of 
the Faddeev amplitude. Both schemes agree for projectile energies
$E_{lab} < 200$ MeV. However, at about 200 MeV deviations occur 
in the cross section peak, as shown in
Fig.~\ref{fig1} for the cross section at $E_{lab} = 197$ MeV, 
where the PW calculation has not sufficiently converged to the 3D calculation. 
The PW calculation shown in Fig.~\ref{fig1}  
takes into account NN angular momenta up to $j = 5$ and $7$, and 3N states for
total 3N angular momenta up to $J = 31/2$,
which is a technical maximum at present.

Since we do not take rescattering terms into account, we nevertheless have
to estimate their effects. At $E_{lab} = 197$ MeV we do this by comparing
the 3D calculation with a full  Faddeev PW calculation, and display the
results in Fig.~\ref{fig2}. We see that the higher order multiple 
scattering contributions lower the cross section \/ and 

\begin{figure}[!h]
\begin{minipage}[t]{60mm}
\begin{center}
\input{figure1.gpl}
\end{center}
\caption{\label{fig1}The spin averaged differential cross section at 
$E_{lab}=197$ MeV, $\theta = 13^0$, based on the Bonn-B potential.} 
\end{minipage}
\hfill
\begin{minipage}[t]{60mm}
are essential for the description of the
analyzing power especially for the small energies of the outgoing
neutron.

\hspace*{1em} Next, we investigate effects of rela\-tivistic kinematics on 
semi-inclusive observables as function of the projectile energy. 
In  Fig.~\ref{fig3} we compare our nonrelativistic 3D calculations with
the corresponding relativistic ones, 
and see that the position as well as the height 
of the quasi-free peak are influenced. Since the position of the quasi-free peak
is solely determined by kinematics, it is satisfying to see that the
use of rela\-tivistic kinematics put the calculated peak at the right 
position with respect to the data. As expected the effect of 
relativistic kinematics increases with increasing projectile energy.
\end{minipage}
\newline
\begin{minipage}[t]{60mm}
\begin{center}
\input{figure2a.gpl}
\end{center}
\end{minipage}
\hfill
\begin{minipage}[t]{60mm}
\begin{center}
\input{figure2b.gpl}
\end{center}
\end{minipage}
\vspace*{0.3cm}
\caption{\label{fig2}The spin averaged differential cross section
  (a) and the analyzing power $A_y$ (b) at $E_{lab}=197$ MeV, $\theta =
  37^0$. The calculations are based on the AV18 potential, the data from
  Ref.~[8].} 
\end{figure}

\begin{figure}[!h]
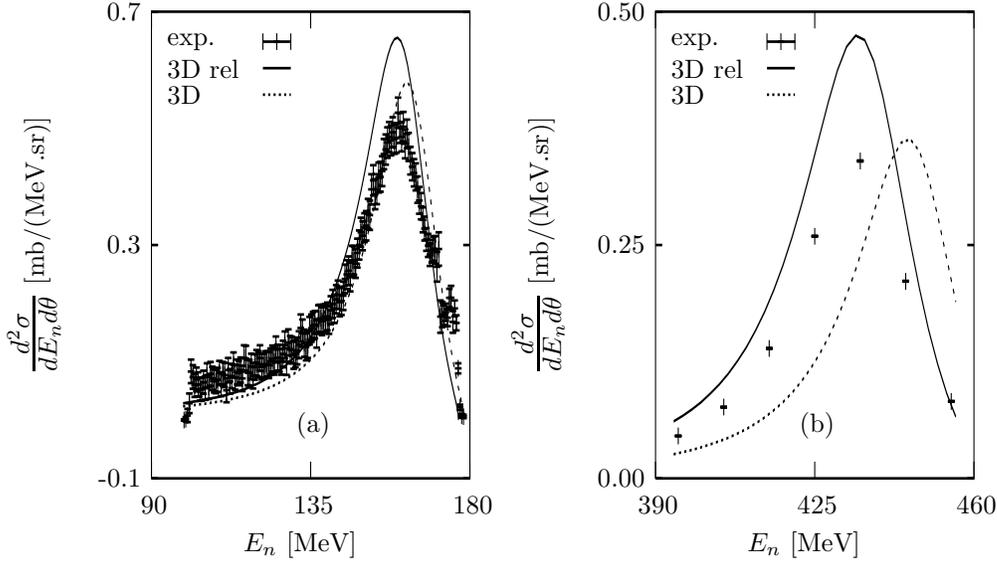

\begin{minipage}[t]{60mm}
\begin{center}
\input{figure3a.gpl}
\end{center}
\end{minipage}
\hfill
\begin{minipage}[t]{60mm}
\begin{center}
\input{figure3b.gpl}
\end{center}
\end{minipage}
\caption{\label{fig3}The spin averaged differential cross section 
at (a) $E_{lab}=197$ MeV, $\theta = 24^0$ and (b) $E_{lab}=495$ MeV, 
$\theta = 18^0$. The calculations are based on the Bonn-B potential, the data from
  Ref.~[8] (a) and Ref.~[6] (b).} 
\end{figure}

\section{Summary and Conclusions}
We calculated the semi-exclusive pd break-up process within the Faddeev 
scheme in the leading order in a multiple scattering expansion in a 3D 
formulation based directly on momentum vectors. This technique has proved to be
a viable alternative to traditional PW decompositions. For energies smaller than
200~MeV our 3D calculations show perfect agreement with PW calculations. At 
energies larger than 200~MeV the inclusion of relativistic kinematics proves
essential to obtain the correct position of the quasi-free peak in the (p,n) charge
exchange reaction.

\end{document}